\documentclass[pdflatex,sn-mathphys-num]{sn-jnl}

\usepackage{graphicx}%
\usepackage{multirow}%
\usepackage{amsmath,amssymb,amsfonts}%
\usepackage{amsthm}%
\usepackage{mathrsfs}%
\usepackage[title]{appendix}%
\usepackage{xcolor}%
\usepackage{textcomp}%
\usepackage{manyfoot}%
\usepackage{booktabs}%
\usepackage{algorithm}%
\usepackage{algorithmicx}%
\usepackage{algpseudocode}%
\usepackage{listings}%
\usepackage{array}
\usepackage{tikz}
\usetikzlibrary{shapes.geometric, arrows.meta, positioning, shadows}

\tikzset{
  flownode/.style = {
    rectangle, rounded corners, draw=black, very thick,
    minimum width=5.5cm, minimum height=1.2cm,
    align=center, fill=white, drop shadow
  },
  arrow/.style = {
    -{Stealth[length=8pt,width=6pt]}, thick
  },
  examplebox/.style = {
    rectangle, draw=gray, rounded corners, inner sep=4pt, fill=gray!5,
    font=\scriptsize, align=center
  }
}

\theoremstyle{thmstyleone}%

\theoremstyle{thmstyletwo}%

\theoremstyle{thmstylethree}%

\title[Article Title]{Institute Disambiguation using Author-Institution Co-Occurrence}

\author*[1]{\fnm{Achal} \sur{Agrawal}}\email{achal.agrawal.1987@gmail.com}

\author[1]{\fnm{Jeet} \sur{Mukherjee}}\email{jeetmukherjee009@gmail.com}

\affil[1]{\orgdiv{PostPub}, \country{India}}

\abstract{In this article we propose a novel method to perform unsupervised clustering of different forms of Institute names. We use only author and affiliation metadata to perform the clustering without any string or pattern matching. After analysing only 50000 articles from Crossref database, we see encouraging results which can be scaled up to provide even better results. We compare our clustering with what a well-known method using string matching does and found that the results were complementary. This can help perform institute disambiguation better when integrated with existing systems, especially to provide aliases for cases where traditional string matching fails. The code of this open-source methodology can be found at: \href{https://github.com/Jeet009/Institute-Disambiguation-using-Author-Institution-Co-Occurrence}{Github}  }

\keywords{Institute Disambiguation, MetaScience, Affiliation Mapping, Publication Statistics, Bibliometrics}

\begin{document}
\maketitle

\clearpage
\section{Introduction}\label{sec1}

Institute Name Disambiguation (IND) has traditionally been a challenging problem. There have been multiple approaches towards solving this problem but most methods incorporate some sort of string comparisons like in Huang \cite{huang2014} and \cite{ROR} .
\vspace{2mm}

In this paper, we depart somewhat from the norm by using the contextual information present in the Author-Institute co-occurrence. The rationale behind it is quite simple. Same author can write the affiliation in different ways. By using authors as the context, we can link the institutes which have the same author. This method might seem somewhat crude as it can potentially have noise due to same author having published with different affiliations. Moreover, two different authors could share the same name and their institutes might again get linked. However, with a sample of just author-institution pairs of 50000 papers and a few elementary filters, we show that this method is remarkably effective. It's unreasonable effectiveness merits further investigation and refinement.
\vspace{2mm}

This method is inspired from word2vec, which found similarity between words occurring in the same context. In this case, we are also using authors as context and assigning similarity based on affiliations which co-occur with the same authors. As a proof of concept, we concentrated on disambiguating Universities in India. This is challenging for most databases as this region usually has less coverage.
\vspace{10mm}
\section{\textbf{Data and Methodology}}

For this study, we considered Crossref metadata of 50000 papers from India published in 2025. Overall there were 150604 unique author names and 77484 unique affiliation names.

We did very basic cleaning of institution names by removing any leading numbers that might have crept in the metadata. The method itself comprises of four minor steps outlined in Fig. \label{fig:steps}.

\begin{itemize}
    \item \textbf{Creation of Author-Affiliation Matrix}: Firstly we convert the unstructured metadata into structured data where Authors are the rows and the affiliations in the papers are the columns. This matrix merely records which author used which affiliation string how many times.
    \item \textbf{Creation of Binary Mask}: Next, we create a binary mask of the Author-Affiliation matrix. This is to ensure that prolific authors do not skew the calculations.
    \item \textbf{Parallelized Calculation of Co-Occurrence Matrix}: This is the most crucial step as it requires a lot of computation. To make it efficient, we simplify the calculation as a mere matrix multiplication and leverage parallelization of TPUs to be able to process 50,000 papers in the free tier.
    \item \textbf{Threshold Based Graph Analysis}: In this step we consider a graph based on the co-occurrence matrix where each string is a node and the weight is the co-occurrence of the two strings. The threshold helps in controlling the precision of the results of the method.
\end{itemize}

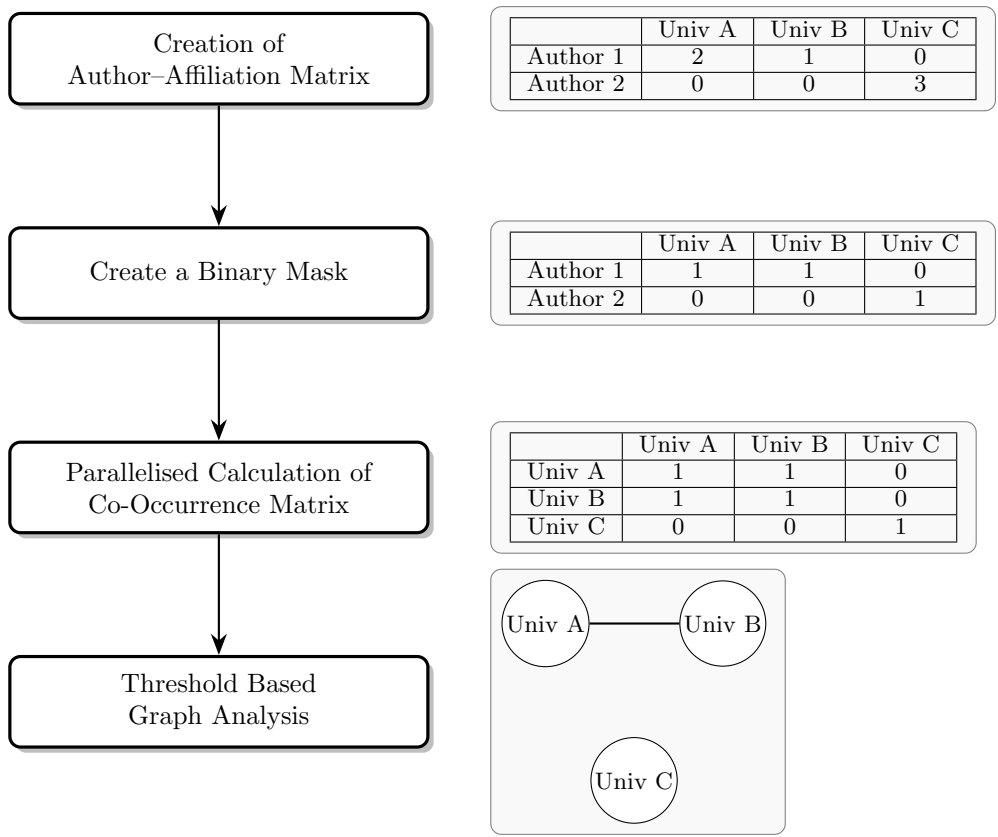
\begin{figure}
    \centering

\begin{tikzpicture}[node distance=16mm and 18mm]

  \node[flownode] (A) {Creation of \\ Author--Affiliation Matrix};
  \node[flownode, below=of A] (B) {Create a Binary Mask};
  \node[flownode, below=of B] (C) {Parallelised Calculation of \\ Co-Occurrence Matrix};
  \node[flownode, below=of C] (D) {Threshold Based \\ Graph Analysis};

  \draw[arrow] (A) -- (B);
  \draw[arrow] (B) -- (C);
  \draw[arrow] (C) -- (D);

  \node[examplebox, right=8mm of A.east, anchor=west] (Aex) {
    \renewcommand{\arraystretch}{1.1}
    \begin{tabular}{|c|c|c|c|}
    \hline
     & Univ A & Univ B & Univ C \\
    \hline
    Author 1 & 2 & 1 & 0 \\
    \hline
    Author 2 & 0 & 0 & 3 \\
    \hline
    \end{tabular}
  };

  \node[examplebox, right=8mm of B.east, anchor=west] (Bex) {
    \renewcommand{\arraystretch}{1.1}
    \begin{tabular}{|c|c|c|c|}
    \hline
     & Univ A & Univ B & Univ C \\
    \hline
    Author 1 & 1 & 1 & 0 \\
    \hline
    Author 2 & 0 & 0 & 1 \\
    \hline
    \end{tabular}
  };

  \node[examplebox, right=8mm of C.east, anchor=west] (Cex) {
    \renewcommand{\arraystretch}{1.1}
    \begin{tabular}{|c|c|c|c|}
    \hline
     & Univ A & Univ B & Univ C \\
    \hline
    Univ A & 1 & 1 & 0 \\
    \hline
    Univ B & 1 & 1 & 0 \\
    \hline
    Univ C & 0 & 0 & 1 \\
    \hline
    \end{tabular}
  };

  \node[examplebox, right=8mm of D.east, anchor=west, inner sep=4pt] (Dex) {
    \begin{tikzpicture}[scale=1.3, every node/.style={circle, draw, fill=white, inner sep=1pt, font=\scriptsize}]
      \node (A1) at (0,0) {Univ A};
      \node (B1) at (1.8,0) {Univ B};
      \node (C1) at (0.9,-1.6) {Univ C};
      \draw[-, thick] (A1) -- (B1);
    \end{tikzpicture}
  };

\end{tikzpicture}

    \caption{Steps involved to perform the clustering. This is different from most Institute Disambiguation Methods as it does not perform any string matching}
    \label{fig:steps}
\end{figure}

\vspace{10mm}
\section{\textbf{Results}}

To cluster the various strings, we perform graph analysis and each connected component corresponds to a cluster. One can check out the results for different thresholds. \url{https://postpub.netlify.app}  .

\subsection{Effect of Threshold}\label{sec2}

\begin{figure}[h!]
\centering
\includegraphics[width=\textwidth]{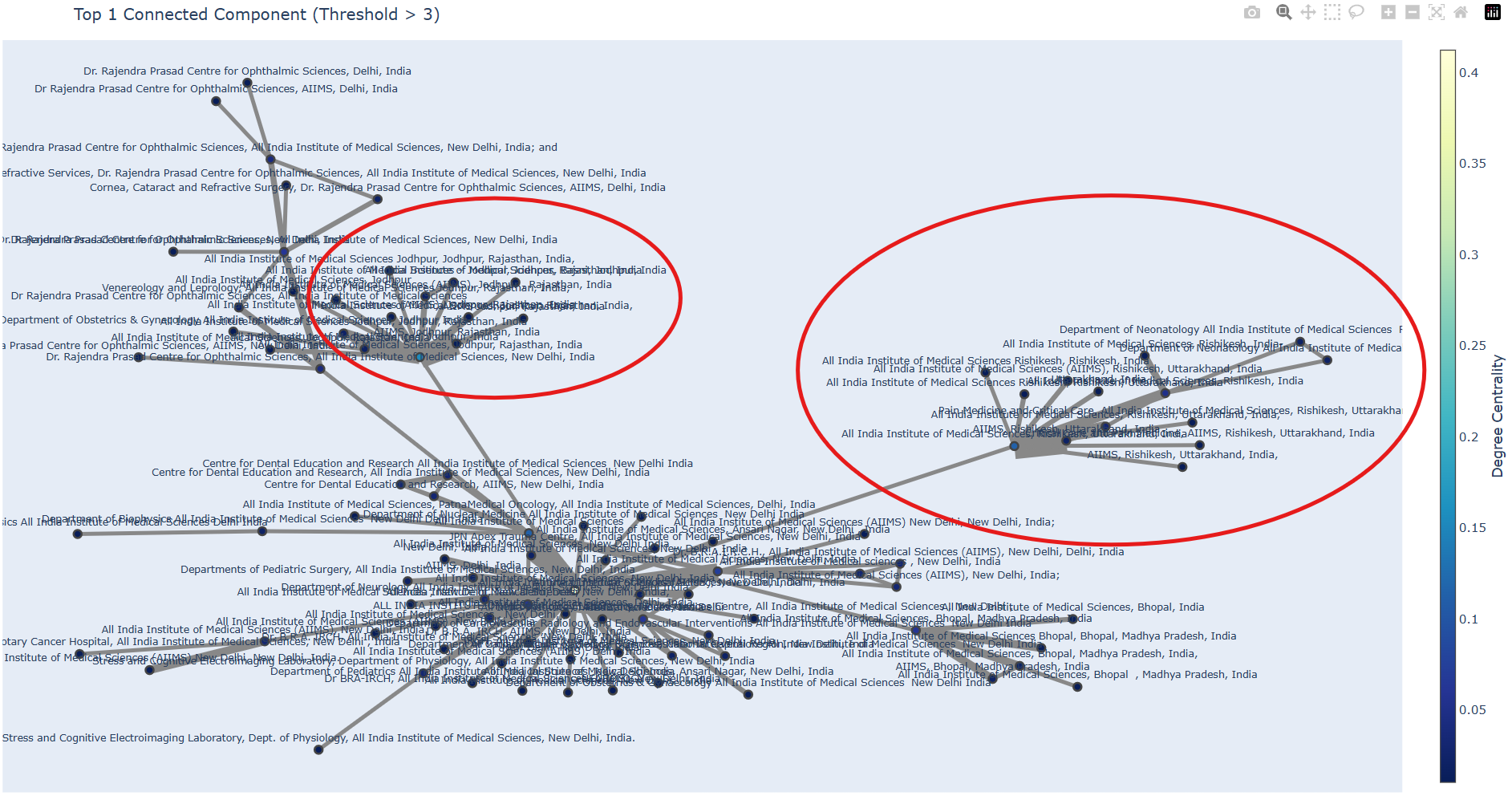}
\caption{Threshold 3 : In this figure, we see that the different forms of writing AIIMS New Delhi neatly cluster together. However, there are a few errors marked in red. These are clusters of different branch of the same university. AIIMS Jodhpur and AIIMS Rishikesh. This can be mitigated by increasing the threshold to 4, it performs better with many more forms of AIIMS New Delhi identified.}\label{fig1}
\end{figure}

\begin{figure}[h!]
\centering
\includegraphics[width=1\textwidth]{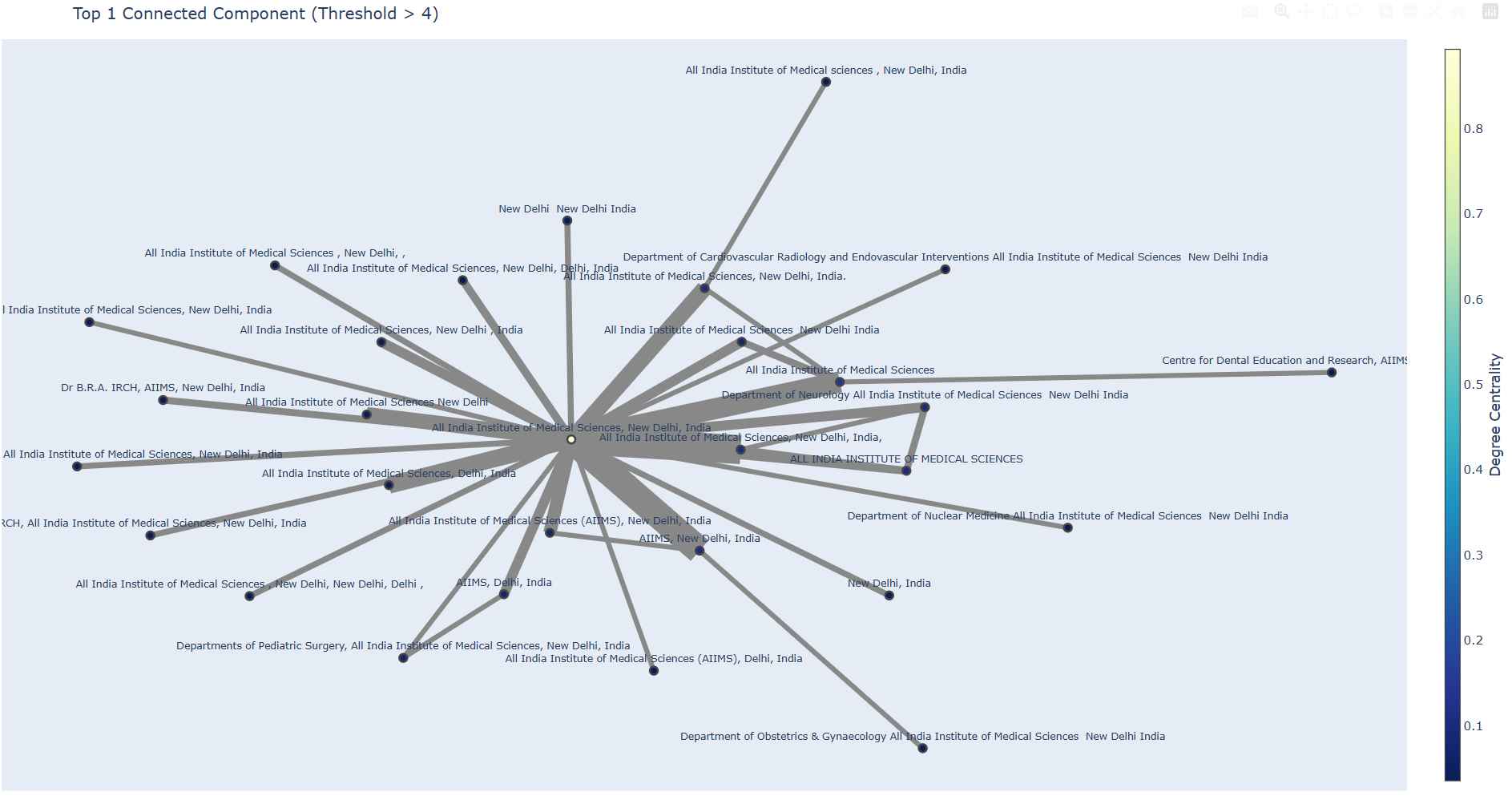}
\caption{Threshold 4 : Higher threshold leads to fewer connected nodes, which is to be expected. However, it also leads to no errors. It shows that threshold forms an important parameter to control the performance of the algorithm. Higher Threshold leads to higher precision but lower recall.}\label{fig2}
\end{figure}

We see that threshold, i.e. the number of times different forms of the university co-occur, is an important parameter while forming the graph. To see the effect of the threshold, In Fig. \ref{fig1} and Fig. \ref{fig2} we plot the biggest cluster in the graph for two different thresholds. In Fig. \ref{fig1} we see that while for a smaller threshold of three, we have many more nodes in the cluster, there are a few errors too. Notably two clusters of different branches of the same university got clustered with it. 
\vspace{2mm}

When the threshold is increased to 4 (Fig. \ref{fig2}), the errors disappear. However the cluster also becomes smaller. This shows the tradeoff between precision and recall. With a much bigger dataset, it will be possible to create more robust graphs.

\subsection{Ability to distinguish between different branches of the same institute}\label{sec2}

\begin{figure}[h]
\centering
\includegraphics[width=1\textwidth]{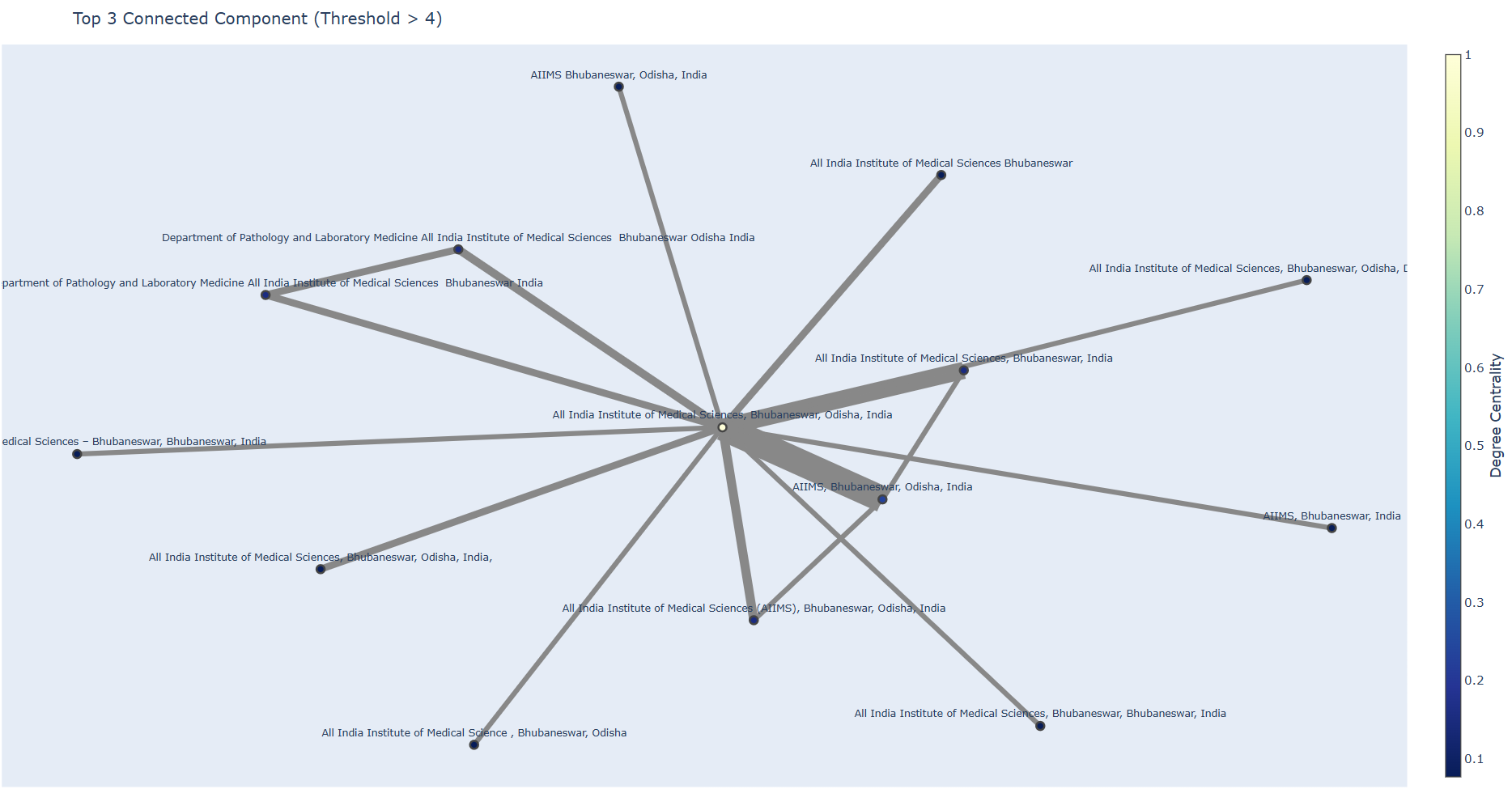}
\caption{Threshold 4 : In this figure, we see the third biggest connected component. It is another branch of AIIMS located in another city. The algorithm manages to distinguish between the two as it is not based on string matching.}\label{fig3}
\end{figure}

A lot of institute disambiguation algorithms fail to differentiate between different branches of the same institute as most of the strings and words match. Using our method however, it automatically manages to put different branches in different clusters as seen in Fig. \ref{fig3}. This is quite useful as doing so manually could be quite tedious.

\subsection{Comparison with String Matching Algorithms}\label{sec2}

One of the more popular algorithm for IND is \cite{huang2014} which uses string matching to create clusters. To check the complementarity of the approaches, we passed the biggest connected component from our algorithm with threshold 3 (99 entries in total). 27 correct strings of the 99 strings were not clustered by this method at all. Even the ones which were clustered were broken into smaller clusters by the algorithm. The biggest cluster created by this algorithm had only 13 components which is much lesser than 28 components for our algorithm with a greater threshold. However, the clusters made by \cite{huang2014} had very high precision with almost no false positive.

It shows that the two approaches are complementary and can benefit from each other. Our method uses no rule based clustering and is almost completely unsupervised. It is thus more robust and applicable to more general settings.
\vspace{10mm}
\section{Future Improvements}
Our analysis is completely unsupervised and does not use any expert rules to simplify the problem. It can however benefit from certain key insights used in string matching methods to make the process even more efficient and robust. Our work can benefit in future from following improvements:

\begin{itemize}
    \item \textbf{Bigger Dataset} - By increasing the size of dataset from 50,000 to a million for a country can dramatically improve results and permit good recall even for higher precision.
    \item \textbf{Incorporating Expert Rules} - We can employ standard well accepted practices in string matching literature to improve the input quality of the algorithm.
    \item \textbf{Do Author Disambiguation before} - Some authors can have homonyms and it can be useful to distinguish between them before creating the Author-Affiliation matrix for better results.

\end{itemize}

The method can help find aliases for affiliations at scale. This in turn can be used by platforms like ROR to perform better using the tested string matching algorithms already developed. It can help reduce a lot of manual work of adding aliases in the ROR database.
\vspace{10mm}
\section{Acknowledgements}

This work is funded by Digital Science Catalyst Grant 2024 for PostPub.


\end{document}